\documentclass{aa}
\usepackage{graphics}
\begin{document}

\def\la{\mathrel{\hbox{\rlap{\hbox{\lower4pt\hbox{$\sim$}}}\hbox{$<$}}}}
\def\ga{\mathrel{\hbox{\rlap{\hbox{\lower4pt\hbox{$\sim$}}}\hbox{$>$}}}}
\def\sq{\hbox{\rlap{$\sqcap$}$\sqcup$}}
\def\arcmin{\hbox{$^\prime$}}
\def\arcsec{\hbox{$^{\prime\prime}$}}
\def\fd{\hbox{$.\!\!^{d}$}}
\def\fh{\hbox{$.\!\!^{h}$}}
\def\fm{\hbox{$.\!\!^{\rm m}$}}
\def\fs{\hbox{$.\!\!^{s}$}}
\def\fdg{\hbox{$.\!\!^\circ$}}
\def\farcm{\hbox{$.\mkern-4mu^\prime$}}
\def\farcs{\hbox{$.\!\!^{\prime\prime}$}}

\newcommand{\etal}{{et al.}\,}      
\newcommand{\eg}{{e.g.},\ }         
\newcommand{\ie}{{i.e.},\ }         
\newcommand{\cf}{{\it cf.},\ }          

\def\deg{{^\circ}}


\title{Extragalactic Large-Scale Structures behind the Southern Milky Way.
\thanks{Tables 1 and 2 are available in electronic format at the CDS
via anonymous ftp to cdsarc.u-strasbg.fr (130.79.128.5) or via
http://cdsweb.u-strasbg.fr/Abstract.html. Based on observations taken at the European Southern
Observatory, La Silla, Chile.}}

\subtitle{IV. Redshifts Obtained with MEFOS}

\author{Patrick A.~Woudt\inst{1} 
\and Ren\'ee C.~Kraan-Korteweg\inst{2} 
\and V\'eronique ~Cayatte\inst{3} 
\and Chantal~Balkowski\inst{3}
\and Paul Felenbok\inst{4}}
\offprints{Patrick A. Woudt,
\email{pwoudt@circinus.ast.uct.ac.za}}

\institute{
Department of Astronomy, University of Cape Town,
Rondebosch 7700, South Africa
\and
Depto. de Astronom{\'{\i}}a, Universidad de Guanajuato, Apartado 
Postal 144, Guanajuato, Gto 36000, Mexico
\and
Obs.~de Paris, GEPI, CNRS and Universit\'e Paris 7, 5 Place Jules Janssen, 92195
Meudon Cedex, France
\and
Obs.~de Paris, LUTH, CNRS and Universit\'e Paris 7, 5 Place Jules Janssen, 92195
Meudon Cedex, France
}

\date{original: 04/06/2003, revised: 27/10/2003}

\abstract{
As part of our efforts to unveil extragalactic large-scale structures behind
the southern Milky Way, we here present redshifts for 764 galaxies in the Hydra/Antlia,
Crux and Great Attractor region ($266^\circ \le \ell \le 338^\circ$, 
$|\,b\,| \la 10^\circ$), obtained with the 
Meudon-ESO Fibre Object Spectrograph (MEFOS) at the 3.6-m telescope of
ESO.
The observations are part of a redshift survey of partially obscured galaxies recorded
in the course of a deep optical galaxy search behind the southern Milky Way (Kraan-Korteweg
2000; Woudt \& Kraan-Korteweg 2001).
A total of 947 galaxies have been observed, a small percentage of the spectra ($N = 109$, 11.5\%) were 
contaminated by foreground stars, and 74 galaxies (7.8\%) were too faint 
to allow a reliable redshift determination.
With MEFOS we obtained spectra down to the faintest galaxies of our optical galaxy survey, and
hence probe large-scale structures out to larger distances ($v \la 30\,000$ km s$^{-1}$) than
our other redshift follow-ups using the 1.9-m telescope at the South African Astronomical
Observatory (Kraan-Korteweg et al.~1995; Fairall et al.~1998; Woudt et al.~1999) and the 64-m
Parkes radio telescope (Kraan-Korteweg et al.~2002).
The most distinct large-scale structures revealed in the southern Zone of Avoidance
are discussed in context to known structures adjacent to the Milky Way.
\keywords { Catalogs -- Surveys -- ISM: dust, extinction -- Galaxies: 
distances \& redshifts -- clusters: individual: 
ACO 3627 -- large-scale structure of Universe}
}

\maketitle

\section{Introduction}

At the same time when significant advances are made in characterising large-scale structures of galaxies in the Universe
through dedicated surveys such as the Sloan Digital Sky Survey (Zehavi et al.~2002) and the 2dF Galaxy Redshift Survey (Colless et al.~2001), 
the study of large-scale structures {\sl behind the Milky Way} has also progressed enormously. 
Dedicated deep optical searches, near infrared all-sky surveys (2MASS and DENIS), HI all-sky surveys
and X-ray surveys, have all resulted in the detection of voids, clusters and superclusters 
at low Galactic latitude (for a review, see Kraan-Korteweg \& Lahav (2000) and references therein). 

We have focussed our efforts on an extended region behind the southern Milky Way (Kraan-Korteweg 2000; Woudt \& Kraan-Korteweg 2001). 
This region of space harbours a large local overdensity of galaxies, the Great Attractor (Dressler et al.~1987;
Lynden-Bell et al.~1988; Kolatt et al.~1995; Tonry et al.~2000). 
Local and complex large-scale structures appear to extend across the Galactic Plane (GP) in this part of the sky
(Kraan-Korteweg et al.~1995 (Paper I); Fairall et al.~1998 (Paper II); 
Woudt et al.~1999 (Paper III)).

Our deep optical galaxy search behind the southern Milky Way is divided into three separate areas:
the Hydra/Antlia region ($266^{\circ} \le \ell \le 296^{\circ}$, $-10^{\circ} \le b \le +8^{\circ}$) 
(Kraan-Korteweg 2000), the Crux region ($289^{\circ} \le \ell \le 318^{\circ}$, $|\, b\, | \le 10^{\circ}$), and 
the Great Attractor region ($316^{\circ} \le \ell \le 338^{\circ}, |\, b\, | \le 10^{\circ}$)
(Woudt \& Kraan-Korteweg 2001). The optical survey in these regions has revealed over 11\,000 previously unrecorded galaxies 
with observed major diameters $\ga 0\farcm2$.  Recently, the survey has been extended towards the Puppis region (at lower
Galactic longitudes), and the Scorpius region (towards the Galactic bulge).

\begin{figure*}
 \resizebox{\hsize}{!}{\includegraphics{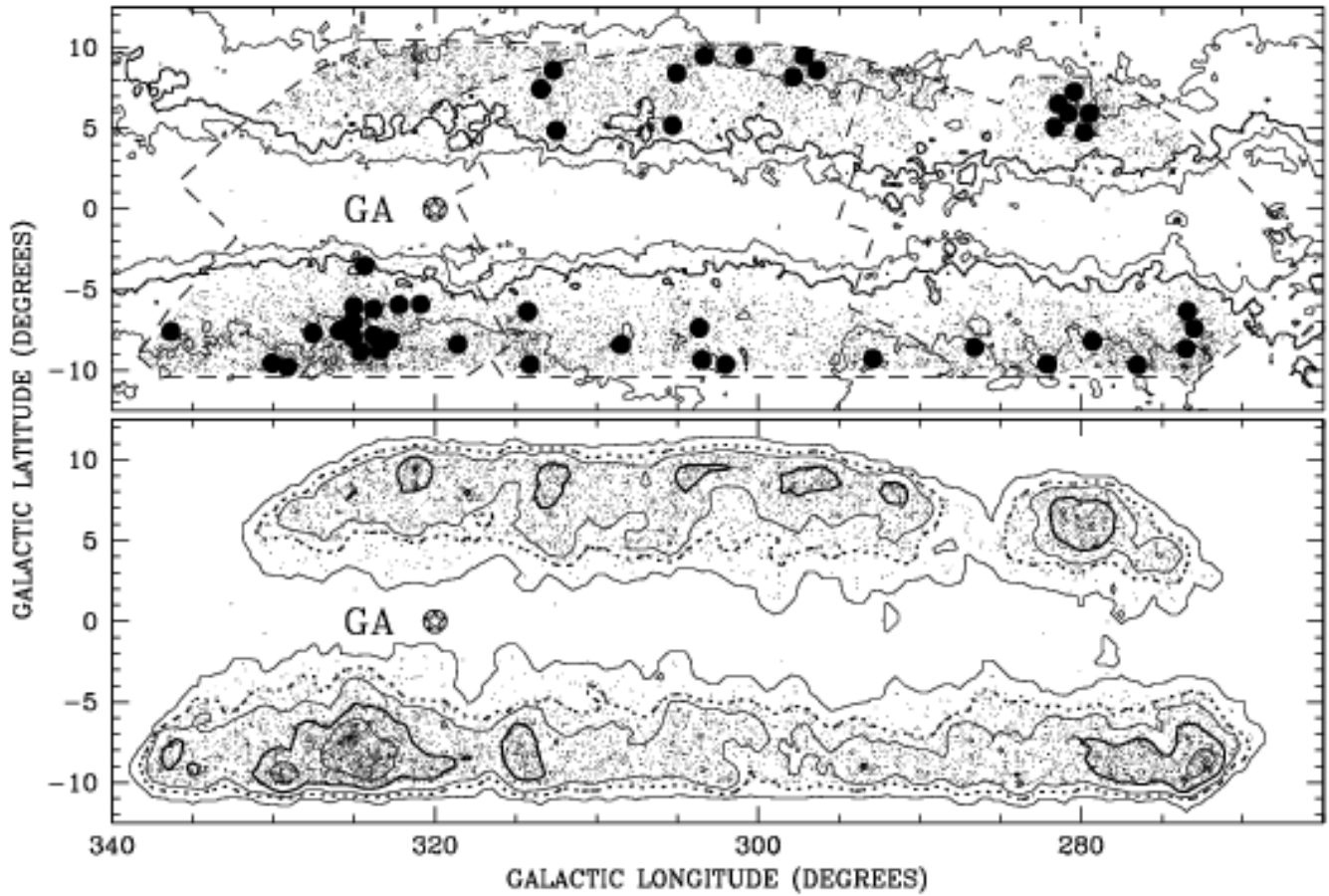}}
\caption{The distribution of galaxies in the Hydra/Antlia, Crux and GA region
(upper panel, from right to left). The dashed line marks these three different survey regions.
The contours are lines of equal Galactic foreground extinction, taken from the Galactic 
reddening maps of Schlegel \etal (1998). The contours correspond to 
$A_{\rm B}$ = 1$^{\rm m}$, 3$^{\rm m}$ (thick line) and 5$^{\rm m}$. In the upper panel, the big dots show the 
location of the 48 observed MEFOS fields. The lower panel shows the galaxy density contours in the
Hydra/Antlia, Crux and GA region. The contours mark 0.5, 5 (dotted line), 10, 25 
(thick line) and 50 galaxies per square degree, respectively. The position of the
peak of the reconstructed mass density field associated with the GA (Kolatt et al.~1995)
is clearly marked.}
\label{meffig1}
\end{figure*}

Following our deep optical galaxy search, we have been engaged
in a redshift survey of these newly found partially obscured galaxies. 
This redshift survey consists of three different, yet complementary 
approaches in tracing the extragalactic large-scale structures. They are:

\begin{itemize}
\item{Optical spectroscopy of individual early-type galaxies with the
1.9-m telescope of the South African Astronomical Observatory (SAAO)
(Papers I -- III).}
\item{H\,I line spectroscopy of low surface brightness spiral and irregular
galaxies with the Parkes 64-m radio telescope (Kraan-Korteweg \etal 2002).}
\item{Low resolution, multi-fibre spectroscopy of galaxies in regions of 
high galaxy-density with the 3.6-m telescope of the European
Southern Observatory (ESO), initially in combination with Optopus,
later with MEFOS (Meudon-ESO Fibre Object Spectrograph).}
\end{itemize}

MEFOS, with its 29 object fibres and large field of view (1-degree diameter), is
optimal for covering the highest density regions in our optical survey.
These multi-fibre observations are an important part of our redshift survey, specifically in
mapping the centre of the GA region and the extended overdensity around ACO 3627. 

In this paper -- paper IV in the series of papers on extragalactic large-scale
structures behind the southern Milky Way -- the results from the MEFOS multi-fibre
spectroscopy in the Hydra/Antlia, Crux and Great Attractor regions are presented.

\section{Observations and data reduction}

MEFOS mounted at the prime focus of the ESO 3.6-m telescope, provided a large field
of view (1 degree in diameter). It had 30 arms that could point within the 20 cm (1 degree) field.
One arm was used for guiding, while the other 29 arms were dedicated to the astronomical objects. Each
arm was equipped with an individual viewing system for accurate setting -- very useful for objects in star
crowded regions -- and carried two spectroscopic fibres, one for the astronomical object, and the other one for 
the sky recording needed for sky subtraction. The spectral fibres intercepted 2.5 arcsec on the sky.

The data presented in this paper were obtained during
10.5 nights in four observing runs between February 1993 and May 1995.
A total of 48 fields were observed in our surveyed region, \ie the Hydra/Antlia,
Crux and Great Attractor (GA) regions. One field was observed twice in order to check the
stability and performance of MEFOS. Also, by observing one field twice, the 
uncertainties in the redshifts obtained following the cross-correlation
procedure (Tonry \& Davies 1979) can be quantified (see Sect.~2.2.).

In all the observing runs, the CCD Tek \#32 was used in combination with grating \#15. The
wavelength coverage spans a range of 3815{\AA} -- 6120{\AA}, with a dispersion of 
170{\AA}/mm and a resolution of $\sim11${\AA}.

Before and after each science frame, an arc exposure was made. In our case, the arcs 
were made with a Helium-Neon lamp. A filter (BG 28) was put in front of the lamp to reduce 
saturation of the bright lines at the red end of the spectrum.
The wavelength-calibration frames had exposure times between one and two minutes, the 
science frames had exposure times of 20 or 30 minutes. Generally, each MEFOS field has been 
observed for 1 hour in total (combining 2 or 3 individual exposures).

The upper panel of Figure~\ref{meffig1} shows the distribution of the observed
MEFOS fields in the Hydra/Antlia, Crux and GA areas. The MEFOS fields
cover the regions of highest galaxy density, with galaxy densities typically in
excess of 25 galaxies per square degree (see lower panel of Fig.~\ref{meffig1}).

\subsection{Data reduction}

All the MEFOS spectra have been reduced with the standard IRAF\footnote{IRAF (Image Reduction
and Analysis Facility) is distributed by the National Optical Astronomy Observatories, which are
operated by the Association of Universities for Research in Astronomy, Inc., under cooperative
agreement with the National Science Foundation.} software package. 
Various command language (cl) programmes were written by V.~Cayatte which greatly streamlined 
the reduction process. This process involved the extraction of the individual spectra,  the wavelength calibration 
using Helium and Neon lamp arc exposures, a correction for the fibre transmission, the sky 
subtraction, the removal of cosmic rays and the determination of the redshift.

Thirteen Helium and Neon emission lines were used for the wavelength calibration. The wavelength solution
to the fit of the arc lines was accurate to better than $\sim 0.3$ {\AA}. A number of fibres
showed a significant offset in the position of the [OI] skyline with respect to its nominal value
of 5577.35 {\AA}, in some cases as much as 1.8 $\pm$ 0.1 {\AA} (Felenbok et al.~1997, but see also Batuski et al.~1999). 
These offsets are systematic and are caused by a misalignment of some of the fibres with respect
to the slit. For those fibres where this shift is larger than 0.5 {\AA}, an inverse wavelength shift
of the average offset (averaged over the entire observing run) to the fibre in question was applied.

The relative fibre transmission coefficient for each fibre is determined from the 
[O\,I] 5577.35{\AA} skyline flux.  For each MEFOS field the measured flux is normalised using the fibre 
with the highest throughput. The transmission might vary from night to night 
due to sky variations or changes in the fibre efficiency (Felenbok \etal 1997). 
The first effect is rather weak below 6000{\AA} (Cuby and Magnoli 1994) and as we find no evidence for 
variations in the overall transmission coefficient (the root-mean-square 
error is 1--3\%) the latter effect is not significant.

However, a clear drop in the fibre transmission coefficients between the 1994 and 1995 data was seen in 6 of the 30 fibres.
In a time span of more than one year some fibres had lost as much as 30\% of their efficiency. 
The likely cause for this drop in efficiency is physical damage to the fibres in question, 
as they are not expected to have any natural decay.

After the fibre transmission corrections were made, the sky spectra within each MEFOS field were examined 
by eye. The sky spectra were averaged to reduce the noise in the sky-subtracted spectrum.

\subsection{Redshift determination}

The IRAF package RVSAO was used to determine the redshifts. Four template stars have been observed; they 
form a template for the cross-correlation of galaxy spectra. The actual cross-correlation of the spectra is done 
with XCSAO within the RVSAO package. After a first cross-correlation (with no a priori imposed velocity limits), 
cross-correlation peaks were checked by eye for each galaxy. Galaxies with an erroneous cross-correlation peak 
of low contrast were either rejected, or, when a clear (unambiguous) peak was seen elsewhere with matching H and 
K Calcium lines, remained in the final list of galaxies. 

The galaxies in this list were then cross-correlated one more time, but now with a strict velocity range imposed. 
For galaxies with redshifts lower than 5000 km s$^{-1}$ this range is $\pm$1000 km s$^{-1}$, for more distant galaxies 
the velocity window ranges from $0.85 \times v_{obs}$ to $1.15 \times v_{obs}$.

Emission line galaxies were treated separately. Their redshifts were determined via the EMSAO task within the 
RVSAO package. The errors in the quoted MEFOS emission line redshifts are conforming to the errors in the SAAO 
data (Papers I--III), i.e., 100 km s$^{-1}$ in case of a single emission line and $100/\sqrt{N}$ in case 
of $N$ emission lines.

\begin{figure}
 \resizebox{\hsize}{!}{\includegraphics{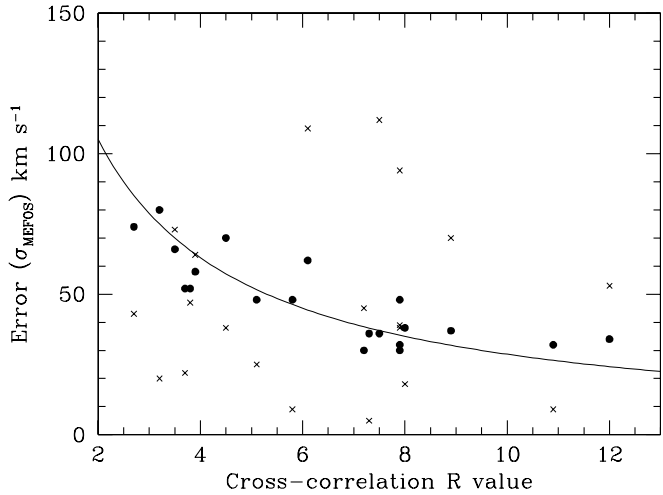}}
\caption{The external error in the MEFOS redshifts, as a function of the $R$-parameter. The crosses represent the
difference in velocity between two independent measurements, the filled circles correspond to the errors used by us. The curve shows
global uncertainties according to the $\sigma_Q$ method, with $Q$ = 315 km s$^{-1}$.}
\label{meffig2}
\end{figure}

The errors in the quoted MEFOS redshifts from the absorption lines are not true `external' errors. They are based on an
internal comparison, rather than a comparison with literature data. One field has been observed twice (field F18) 
resulting in 16 galaxies having two independently measured MEFOS redshifts. A further 4 galaxies were found on overlapping fields. 
For these 20 galaxies, the difference between the two redshift measurements is plotted (as crosses) in 
Figure~\ref{meffig2} against the Tonry and Davies (1979) $R$-parameter. 

The $\sigma_Q$ method (Hill \& Oegerle 1993, Pinkney \etal 1993) is often used to infer an error 
($\sigma_Q = {Q/(R+1)}$) on the basis of the observed R-parameter (contrast factor). Based on the 20 overlapping galaxies
displayed in Figure~\ref{meffig2}, ${Q_{\rm MEFOS}} = 315$ km s$^{-1}$ corresponds best to the uncertainties in 
our MEFOS data, i.e., the solid line drawn in Figure~\ref{meffig2}). This corresponds reasonably well with the $Q$-value
obtained by Batuski et al.~(1999) from their MEFOS observations.

The solid dots in Figure~\ref{meffig2} are the errors as quoted in our redshift catalogue.
These errors correspond to twice the standard deviation of the $R$-weighted average velocity. 
These errors, rather than $\sigma_Q$, were used in our redshift catalogue. They are, as can be 
seen in Figure~\ref{meffig2}, entirely consistent with $Q = 315$ km s$^{-1}$.

\subsection{The redshift catalogue}

The 764 redshifts obtained with MEFOS are presented in Table 1\footnote{Table 1 is only available in electronic format
at the CDS via anonymous ftp to cdsarc.u-strasbg.fr (130.79.128.5) or via
http://cdsweb.u-strasbg.fr/Abstract.html.}. The entries in Table 1 are as follows:

\begin{description}
\item{{\bf Column 1 and 2}: Identification of the galaxy as given in Kraan-Korteweg (2000) -- prefix `RKK' -- and Woudt \& Kraan-Korteweg (2001) -- prefix `WKK' --, and Lauberts identification (Lauberts 1982).}
\item{{\bf Column 3 and 4}: Right Ascension and Declination (2000.0). The positions were measured with the Optronics
machine at ESO in Garching and have an accuracy of about 1 arcsec.}
\item{{\bf Column 5 and 6}: Galactic longitude $\ell$ and latitude $b$.}
\item{{\bf Column 7}: Major and minor axes (in arcsec). These diameters are measured approximately to the isophote
of 24.5 mag arcsec$^{-2}$ and have a scatter of $\sigma \approx 4''$.}
\item{{\bf Column 8}: Apparent magnitude $B_{\rm 25}$. The magnitudes are estimates from the film copies
of the SRC IIIaJ Survey based on the above given diameters and an estimate of the average surface
brightness of the galaxy.}
\item{{\bf Column 9}: Morphological type. The morphological types are coded similarly to the precepts of the Second
Reference Catalogue (de Vaucouleurs et al.~1976). Due to the varying foreground extinction a homogeneous and detailed
type classification could not always be accomplished and some codes were added: In the first column F for E/S0 was added
to the normal designations of E, L, S and I. In the fourth column the subtypes E, M and L are introduced next to the general
subtypes 0 to 9. They stand for early spiral (S0/a -- Sab), middle spiral (Sb -- Sd) and late spiral (Sdm -- Im). The
cruder subtypes are a direct indication of the fewer details visible in the obscured galaxy image. The question
mark at the end marks uncertainty of the main type, the colon marks uncertainty in the subtype.}
\item{{\bf Column 10:} Heliocentric velocity ($cz$) and error as derived from the absorption features. The errors may
appear large as they are estimated external errors, and not internal errors (see Sect.~2.2). The square brackets 
indicate a tentative redshift.}
\item{{\bf Column 11:} Heliocentric velocity and error measured from the emission lines (identified in Column 12) when
present. The square brackets indicate a tentative redshift.}
\item{{\bf Column 12:} Code for the emission lines identified in the respective spectra: \hfill \phantom{a}

\smallskip
\begin{tabular}{ccccc}
 1 & 2 & 3 & 4 & 5 \\
$[$OII$]$ & H$\gamma$ & H$\beta$ & [OIII] & [OIII] \\
3727 & 4340 & 4861 & 4959 & 5007 \\
\end{tabular}
}

\item{{\bf Column 13:} Code for additional remarks: 

\smallskip
* -- Redshift is also available in the literature (NASA Extragalactic Database (NED), May 2003).

\smallskip
1 -- RKK 1808: The redshift measured with MEFOS is in slight disagreement with the value
quoted in the literature ($v = 5571 \pm 150$ km s$^{-1}$, Paper I). Preference is given to the value
quoted here.

\smallskip
2 -- RKK 2313: The redshift measured with MEFOS is in slight disagreement with the (tentative) value
quoted in the literature ($v = 5741 \pm 300$ km s$^{-1}$, Paper I). Preference is given to the value
quoted here.

\smallskip
3 -- WKK 662: The redshift measured with MEFOS is in slight disagreement with the (tentative) value
quoted in the literature ($v = 6526 \pm 231$ km s$^{-1}$, Paper II). Preference is given to the value
quoted here.

\smallskip
4 -- WKK 708: The redshift measured with MEFOS is in disagreement with the (tentative) value
quoted in the literature ($v = 1326 \pm 243$ km s$^{-1}$, Paper II). The clear emission lines in the MEFOS
spectra support the rejection of the literature value.

\smallskip
5 -- WKK 1883: The redshift quoted in the literature ($v = 6801 \pm 70$ km s$^{-1}$, Visvanathan \& Yamada 1996) 
is in disagreement with the redshift determined from the MEFOS spectra. The latter shows a clear range of absorption lines 
and its redshift determination is unambiguous. Preference is given to the MEFOS value.

\smallskip
6 -- WKK 5094: The redshift measured with MEFOS is in disagreement with the (tentative) value
quoted in the literature ($v = 17956 \pm 250$ km s$^{-1}$, Paper III). Preference is given to the value quoted
here.

\smallskip
7 -- WKK 5366: The redshift quoted in the literature ($v = 2059 \pm 10$ km s$^{-1}$, Juraszek et al.~2000) 
is imcompatible with the MEFOS redshift. There is a galaxy (WKK 5365) $\sim 3.5'$ from WKK 5366, and the literature
value (H\,I line spectroscopy) could have suffered from source confusion.

\smallskip
8 -- WKK 5416: The redshift quoted in the literature ($v = 12403 \pm 70$ km s$^{-1}$, Visvanathan \& Yamada 1996)
is in disagreement with the redshift determined from the MEFOS spectra. The latter shows a clear range of absorption lines 
and its redshift determination is unambiguous. Preference is given to the value quoted here.
}
\end{description}

In Table 2\footnote{Table 2 is only available in electronic format
at the CDS via anonymous ftp to cdsarc.u-strasbg.fr (130.79.128.5) or via
http://cdsweb.u-strasbg.fr/Abstract.html.}, 
we present the 183 galaxies observed with MEFOS, but for which no redshift could be obtained due to
the presence of superimposed foreground stars on the galaxy, or due to the low signal-to-noise of the resulting spectra.
The entries in Table 2 are as follows:

\begin{description}
\item{{\bf Column 1 -- 8}: As in Table 1.}

\item{{\bf Column 9}: As in Table 1. In addition, a question mark prior to the galaxy classification indicates
uncertainty about the galaxian nature of the candidate.}

\item{{\bf Column 10}: Remarks about the quality of the spectrum.}
\end{description}

\subsection{Comparison to other measurements}

There are 71 galaxies in Table 1 for which a redshift has been obtained previously (NED, May 2003). 
For 8 of them, the value quoted
here is in disagreement with the literature data (see Column 13 of Table 1), but for the remaining 63 galaxies there is good
agreement. We find that

\begin{displaymath}
<v_{\rm MEFOS} - v_{\rm literature}> = -8 \pm 18 {\rm \ km \,\, s^{-1}}.
\end{displaymath}  

\begin{figure}
 \resizebox{\hsize}{!}{\includegraphics{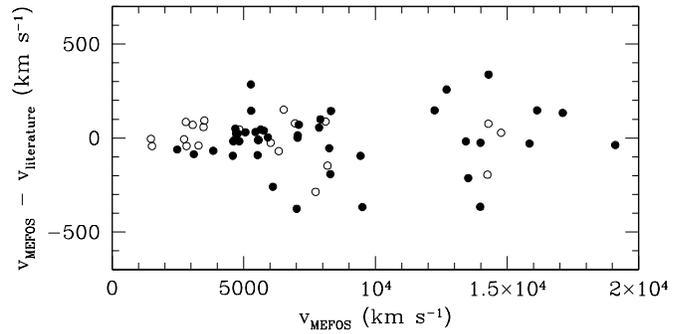}}
\caption{The velocity difference ($v_{\rm MEFOS} - v_{\rm literature}$) as a function of $v_{\rm MEFOS}$. The filled circles correspond
to the comparison with the SAAO data (Papers I -- III), the open circles correspond to the remainder of the literature sample.}
\label{meffig3}
\end{figure}

If the overlapping sample is restricted to measurements made by us at the SAAO (Papers I -- III), the 
agreement again is good, although the standard deviation is somewhat larger 

\begin{displaymath}
<v_{\rm MEFOS} - v_{\rm SAAO}> = -9 \pm 24 {\rm \ km \,\, s^{-1}},
\end{displaymath}  

based on 42 galaxies in common. The larger standard deviation is due to the increasing uncertainty in the SAAO spectra 
of the more distant galaxies. Fig.~\ref{meffig3} shows the general good agreement between the MEFOS redshifts and
the literature values.

\begin{figure*}
 \resizebox{\hsize}{!}{\includegraphics{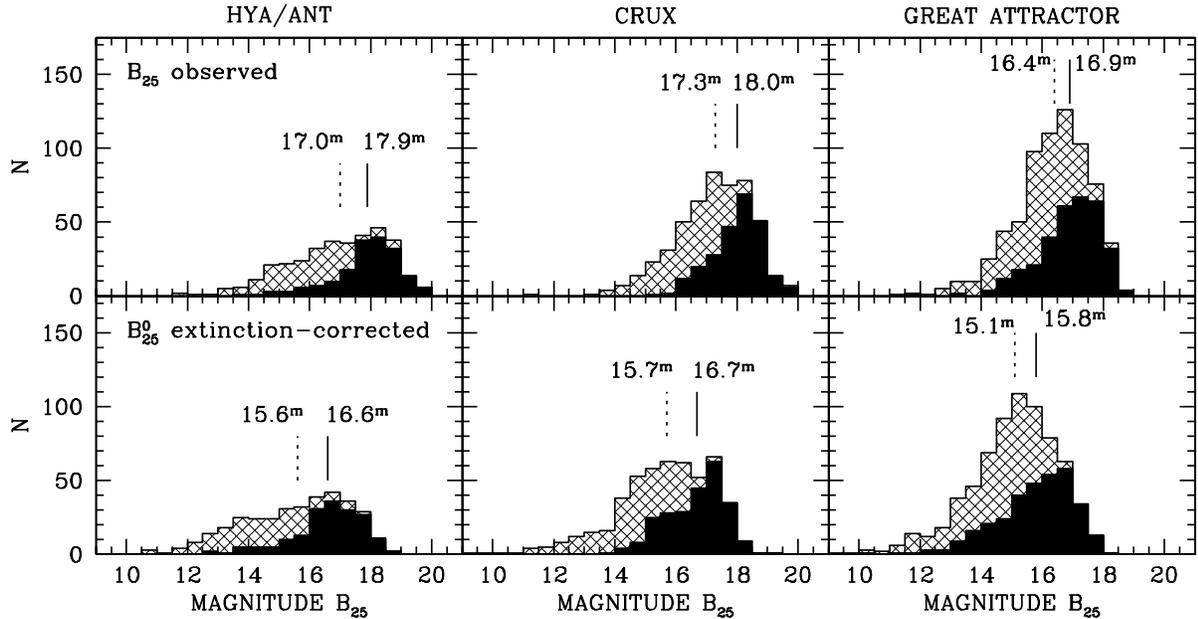}}
\caption{The upper row of panels show the magnitude ($B_{25}$) distribution of galaxies in
the combined (literature + MEFOS) sample (cross-hatched histogram) and MEFOS sample alone (black histogram), for the
Hydra/Antlia (left), Crux (middle) and GA region (right), respectively. The lower row of panels show the extinction-corrected
magnitude ($B_{25}^0$) distribution. The symbols are the same as the upper row of panels. The mean values for the combined, and MEFOS
sample are indicated by the dashed, and solid vertical line for the combined (literature + MEFOS), and MEFOS sample, respectively.}
\label{mefoscomp}
\end{figure*}

\section{Coverage and completeness}

In the following two sections, only the galaxies with a reliable redshift ($N = 754$) will be used in the discussion and 
plots; the galaxies for which we obtained a tentative redshift ($N = 10$) will not be included.

\subsection{The Hydra/Antlia region}

The Hydra/Antlia region contains 3279 galaxy candidates (Kraan-Korteweg 2000). 183 Galaxies in this region have a previously determined
redshift (NED, and references therein). Most of these redshifts were determined in the course of our redshift survey 
(Kraan-Korteweg et al.~1995; Kraan-Korteweg et al.~2002). 

The MEFOS observations have resulted in 185 redshifts in the Hydra/Antlia region (see Table 1), of which 7 are tentative 
measurements. For eighteen galaxies, a redshift had been published previously, so 160 new redshifts in the Hydra/Antlia region 
are presented here. A total of 343 galaxies (literature + MEFOS) in the Hydra/Antlia region
(= 10.5\%) now have a reliable redshift. This percentage is somewhat lower than that in the Crux and GA regions 
(see Sect. 3.2. and 3.3.), but additional spectra have been obtained for 500 galaxies with Optopus (Kraan-Korteweg et al.~1994),
and will be presented elsewhere.

The upper-left panel of Fig.~\ref{mefoscomp} shows the magnitude distribution ($B_{25}$) of all the 343 galaxies with a reliable redshift
(cross-hatched histogram). The filled histogram shows the magnitude distribution of the 178 galaxies with a reliable MEFOS redshift.
It is clear that the MEFOS observations trace the fainter end of the magnitude distribution of the galaxies in the Hydra/Antlia region.
The average $B_{25}$ of the MEFOS sample is 17$\fm$9, only slightly brighter than the overall mean magnitude of the galaxies in
the Hydra/Antlia catalogue, 18$\fm$2 (Kraan-Korteweg 2000). 

The mean extinction-corrected magnitude of the MEFOS sample is $<$$B_{25}^{0}$$>$ = 16$\fm$6, compared to $<$$B_{25}^{0}$$>$ = 15$\fm$6 for the entire
redshift sample (literature and MEFOS). We have used the Galactic reddening maps of Schlegel et al.~(1998) for the extinction correction; see Woudt \&
Kraan-Korteweg (2001) for full details of the extinction correction procedure.

\subsection{The Crux region}

In the Crux region, 3759 galaxy candidates have been identified (Woudt \& Kraan-Korteweg 2001), and 270 of them have a reliable
redshift reported in the literature. 
The MEFOS observations have resulted in 251 galaxy redshifts in the Crux region (see Table 1), 
of which 2 were tentative measurements. Fifteen galaxies of the MEFOS sample 
had a previous redshift measurement, so 234 new reliable redshifts have been obtained. This increases the total number of galaxies in the Crux
region with a redshift to 504 (13.4\%).

As before in the Hydra/Antlia region, the MEFOS sample contributes predominantly to the faint end of the galaxy distribution
(upper-middle panel of Fig.~\ref{mefoscomp}). The mean magnitude of the 249 galaxies MEFOS sample is $<$$B_{25}$$>$ = 18$\fm$0, again only slightly
brighter than the overall mean magnitude of all the galaxies in the Crux region ($<$$B_{25}$$>$ = 18$\fm$2, Woudt \& Kraan-Korteweg 2001). 

\subsection{The Great Attractor region}
 
4423 Galaxy candidates have been identified in the Great Attractor region (Woudt \& Kraan-Korteweg 2001), 413 of them have a reliable
redshift (NED, May 2003). The Norma cluster (ACO 3627, Abell et al.~1989) is the region of highest galaxy density in the GA region,
and as a result a large number of the MEFOS fields were centred on the Norma cluster (see also the lower panel of Fig.~\ref{meffig1}).
We have obtained 328 redshifts in the GA region, one of which is a tentative measurement. Of the remaining 327 galaxies, 38 had a 
previously published redshift, therefore the MEFOS spectroscopy has resulted in 289 new galaxy redshifts in the GA region. A total of 702 galaxies
(15.9\%) of all the galaxies in the GA region now have a redshift (literature + MEFOS).

The mean brightness of the 327 galaxies in the MEFOS sample is $<$$B_{25}$$>$ = 16$\fm$9, about 1 magnitude brighter than the MEFOS sample in
the Hydra/Antlia and Crux regions. This is due to the presence of a nearby overdensity -- the rich Norma cluster (Kraan-Korteweg
et al.~1996) -- in the GA region, whereas most of the regions of high galaxy density in the Hydra/Antlia and Crux region correspond
to more distant clusters (see Sect.~4).

\subsubsection{The Norma cluster}

Within the Abell radius of the Norma cluster (defined as $3 \, h_{50}^{-1}$ Mpc), 603 galaxies are present in our galaxy catalogue 
(Woudt \& Kraan-Korteweg 2001). We have now obtained redshifts for 265 of them (= 44\%) -- 130 of these were obtained with MEFOS.
Of the 265 galaxies with a reliable redshift, 214 are comfirmed cluster members. The others are clear foreground/background galaxies. Prior to our
redshift survey, only 17 redshifts were known within the Abell radius of the Norma cluster.

For galaxies with $B_{25}^0 \le 15\fm5$, 80\% now have a reliable redshift. Most of the missing bright galaxies are low surface brightness
spirals, which have been observed with the Parkes 64-m radio telescope and will be presented elsewhere.


A detailed dynamical analysis of the Norma cluster will be presented elsewhere (including recently obtained 2dF spectroscopy of the Norma
cluster). It is clear, however, that the MEFOS observations form an essential basis for this analysis.

\section {Identification of large-scale structures}

\begin{figure}
 \resizebox{\hsize}{!}{\includegraphics{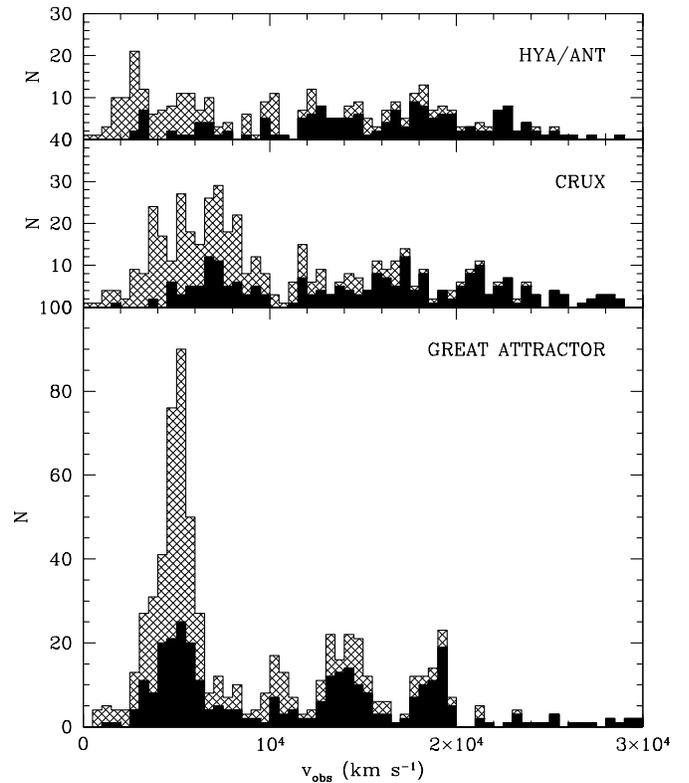}}
\caption{Velocity histogram of the galaxies in the Hydra/Antlia region (upper panel), the Crux region (middle panel) and the
GA region (lower panel). The dark shaded area correspond to the heliocentric velocities obtained with MEFOS, the 
cross-hatched region shows all the heliocentric velocities available to date.}
\label{mefvelhist}
\end{figure}

\subsection{Velocity distribution}

The velocity histograms in Fig.~\ref{mefvelhist} show that distinctly different large-scale structures are present in the
Hydra/Antlia, Crux and GA region, respectively.

The Hydra/Antlia region shows a strong peak at $v \sim 2750$ km s$^{-1}$, which corresponds to the extension of
the Hydra/Antlia supercluster in to the Zone of Avoidance (Paper I; Kraan-Korteweg 2000). In addition, an overdensity at $\sim$ 6000 km s$^{-1}$
is present, associated with the Vela overdensity (Paper I). 
The MEFOS spectra have not added much information to the nearby
structures ($< 10\,000$ km s$^{-1}$), but instead have unveiled more distant galaxies. Most of the redshifts of galaxies
beyond 10\,000 km s$^{-1}$ have come from the MEFOS spectroscopy. In the Hydra/Antlia region, there are 16 galaxies with heliocentric
velocities in excess of 30\,000 km s$^{-1}$ (not shown in Fig.~\ref{mefvelhist}). These are all obtained with MEFOS.

In the Crux region, a broad concentration of galaxies around $3500 < v < 8500$ km s$^{-1}$ is present (middle panel of Fig.~\ref{mefvelhist}).
This broad concentration -- a combination of the Centaurus-Crux cluster (Woudt 1998, also known as CIZA J1324.7-5736 (Ebeling et al.~2002))
and the large-scale overdensity between the Norma cluster and the Vela cluster (see also panel C of Fig.~\ref{mefsky}) --
is very different from the distinct peaks at 2750 km s$^{-1}$ and 5000 km s$^{-1}$, in the Hydra/Antlia and Great
Attractor region, respectively. Again, the MEFOS observations trace the more distant galaxies more efficiently compared to the
earlier SAAO data (Papers I -- III). In the Crux region, 32 galaxies are located beyond 30\,000 km s$^{-1}$. Again, all these 
redshifts were obtained with MEFOS.

The GA region is dominated by four distinct peaks, the most prominent of which is associated with the Norma
cluster and its surrounding great wall-like structure at 4848 $\pm$ 61 km s$^{-1}$ (Kraan-Korteweg et al.~1996; Woudt 1998). 
The other peaks are at 10\,406 $\pm$ 82 km s$^{-1}$, 14\,050 $\pm$ 84 km s$^{-1}$ and 18\,689 $\pm$ 81 km s$^{-1}$.  The peak at 14\,050 km s$^{-1}$ belongs to the
Ara cluster (Woudt 1998; Paper III; Ebeling et al.~2002). Together with the adjacent Triangulum-Australis cluster (McHardy et al. 1981),
the Ara cluster is part of a larger overdensity referred to as a 'Greater Attractor behind the Great Attractor' by 
Saunders et al.~(2000). Saunders et al. find evidence for such an overdensity at $\sim$ 12\,500 km s$^{-1}$ and ($\ell, b$) 
$\approx$ $(326\deg, -3\deg)$ from the reconstructed IRAS galaxy density field.  In the GA region there are
18 galaxies more distant than 30\,000 km s$^{-1}$, all derived from MEFOS observations.

\subsection{Sky projection; local cosmography}

\begin{figure*}
 \resizebox{\hsize}{!}{\includegraphics{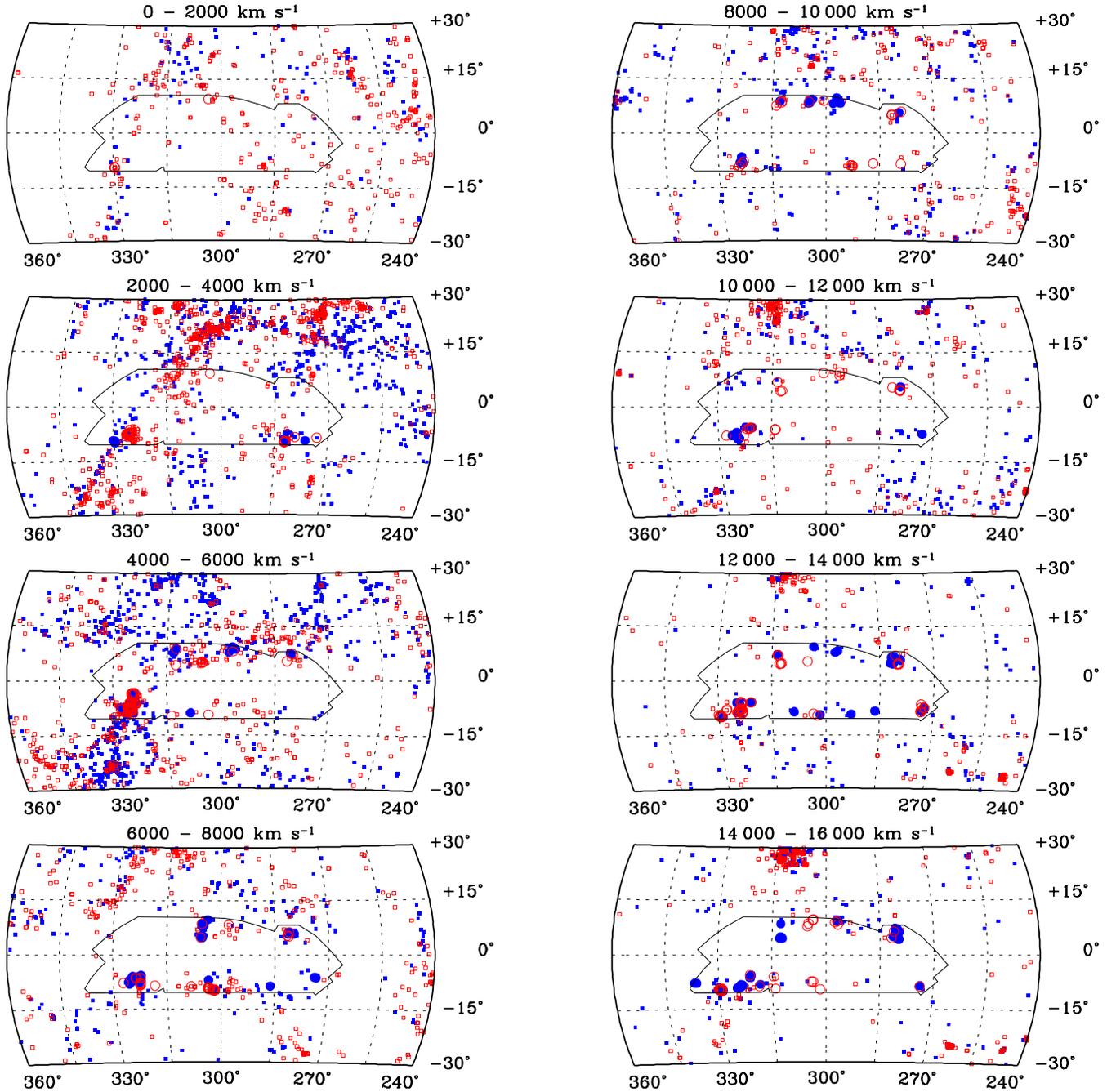}}
\caption{An Aitoff projection of galaxies in the range of $240\deg \le \ell < 360\deg$ and $-30\deg < b < +30\deg$, in heliocentric
velocity intervals of $\Delta v = 2000$ km s$^{-1}$. These graphs
are centred on $\ell = 300\deg$ and $b = 0\deg$ with decreasing longitudes
towards the right. Our search area is marked and the velocity range is indicated on the top of each panel.
Within each panel, the filled symbols (small squares: NED, big circles: our MEFOS data) show the nearer 1000 km s$^{-1}$ slice,
the open symbols (again, small squares: NED, big circles: our MEFOS data) show the more distant 1000 km s$^{-1}$ slice.}
\label{mefsky}
\end{figure*}

In Fig.~\ref{mefsky}, all the galaxies out to 16\,000 km s$^{-1}$ are shown in an Aitoff projection of a limited area of the sky
($240\deg \le \ell < 360\deg$, $-30\deg < b < +30\deg$).
They are separated in eight individual slices, each 2000 km s$^{-1}$ thick, but within each panel the different symbols mark a subdivision:
the filled symbols show the `nearer' 1000 km s$^{-1}$ (e.g. 0 -- 1000 km s$^{-1}$ in the top-left panel), the open symbols
the `further' 1000 km s$^{-1}$ (e.g. 1000 -- 2000 km s$^{-1}$ in the top-left panel). 
The data shown are literature values (small squares: NED, May 2003)
combined with the MEFOS heliocentric velocities (big circles). The Aitoff projections are centred on $\ell = 300\deg$ and $b = 0\deg$.

Here, we will discuss the unveiled large-scale structures behind the southern Milky Way out to 16\,000 km s$^{-1}$,
in relation to the structures visible away from the Galactic Plane (GP).

{$v \le 2000$ km s$^{-1}$:} 
The Supergalactic Plane (SGP) is visible as a concentration
of galaxies starting from the Virgo cluster (located outside this panel at $(\ell, b)$ = ($280\deg, +74\deg$)), 
crossing the GP at $\ell = 325\deg$, and continuing down towards the south
Galactic pole. Also, the Fornax wall is discernable in the 1000 -- 2000 km s$^{-1}$ range 
as an overdensity of galaxies starting from the Fornax cluster at 
($\ell, b$) = ($237\deg, -54\deg$) towards (and crossing) the GP at $\ell = 290\deg$.
The GA overdensity is dominant in the next two slices.

{$2000 < v \le 4000$ km s$^{-1}$:} The vertical band of galaxies associated with the SGP in the previous
slice is very pronounced in this redshift slice (and the next one); this is the Centaurus wall, a very extended structure (both on 
the sky, as in depth), incorporation both the Centaurus cluster ($\ell, b$) = ($302\deg, +22\deg$) and the Norma cluster 
($\ell, b$) = ($325\deg, -7\deg$). The Norma cluster is visible in this panel, but is most pronounced in the range 4000 -- 6000 km s$^{-1}$.
A wall-like structure is also visible crossing the GP at $\ell = 240\deg$. This is the Hydra wall, encompassing the Hydra 
cluster ($\ell, b$) = ($270\deg, +27\deg$) and the Puppis cluster ($\ell, b$) = ($240\deg, 0\deg$), and 
continuing towards ($\ell, b$) = ($210\deg, -30\deg$).

{$4000 < v \le 6000$ km s$^{-1}$:} A large structure of galaxies is seen below the GP extending from $(\ell, b$) $\approx$ ($30\deg,
-60\deg$) -- located outside the boundary of this plot -- to the Norma cluster. 
This structure, which is part of the Norma supercluster, is under a slight angle with respect
to the Centaurus wall in the previous two slices. Here, the Norma supercluster and the Centaurus wall possibly intersect.
Also, parts of the Norma supercluster can be seen (in the more distant 5000 -- 6000 km s$^{-1}$ range -- 
north of the GP around $\ell = 280\deg - 315\deg$, incorporating the
Vela cluster (Kraan-Korteweg \& Woudt 1994) and the Centaurus-Crux cluster (Woudt 1998).  From X-ray observations,
the latter (CIZA\,J1324.7-5736) appears possibly as massive as the Norma cluster (Ebeling et al.~2002). 

{$6000 < v \le 8000$ km s$^{-1}$:} The finger-of-god effect for both the Norma cluster and CIZA J1324.7-5736 cluster 
appear in this panel. In this velocity range, few, or none, extented structures are seen in the southern
sky. 

{$8000 < v \le 10\,000$ km s$^{-1}$:} There is a slight overdensity north of the Galactic plane at $\ell \approx 295\deg$
in the nearer range (filled circles), but no distinct clusters are visible in our surveyed region. Note, however, the Ophiuchus cluster
(Hasegawa et al.~2000) at the edge of the plot ($\ell, b, v$) $\approx$ ($360\deg, +9\deg$, 8500 km s$^{-1}$).

{$10\,000 < v \le 12\,000$ km s$^{-1}$:} The Shapley region comes into view at ($\ell, b$) $\approx$ ($315\deg, +25\deg$).
Behind the Norma cluster ($\ell \approx 325\deg$) an overdensity is visible (continuing into the next panel), corresponding to the peak in
Fig.~\ref{mefvelhist} at 10\,500 km s$^{-1}$.

\begin{figure*}
 \resizebox{\hsize}{!}{\includegraphics{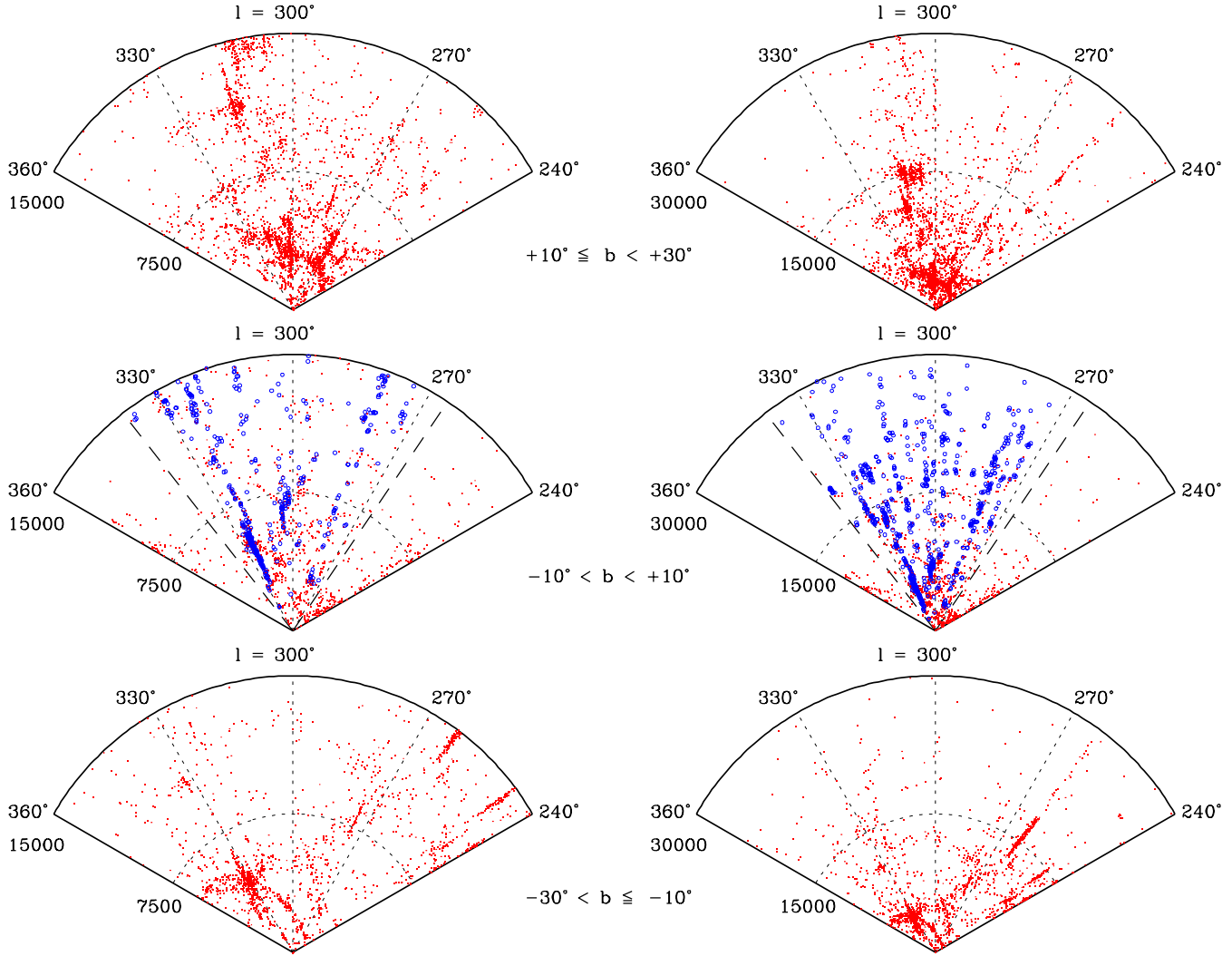}}
\caption{Redshift slices out to $v < 15\,000$ km s$^{-1}$ 
(left column) and $v < 30\,000$ km s$^{-1}$ (right column)
for the longitude range $240\deg \le \ell < 360\deg$. The
top panels display the structures above the Galactic Plane ($+10\deg \le b < +30\deg$),
the middle panels show the structures behind the Galactic Plane ($-10\deg < b < +10 \deg$),
and the lower panels show the large-scale structures at $-30\deg < b \le -10\deg$.
The dashed lines in the middle panels show the demarcation of our survey area.
The data shown are literature values (small dots: NED, May 2003)
combined with the MEFOS heliocentric velocities (open circles).}
\label{mefpiel}
\end{figure*}

{$12\,000 < v \le 16\,000$ km s$^{-1}$:} In these more distant slices, the Shapley region dominates the galaxy distribution
above the GP, and within our surveyed region various clusters are visible (see also the pie diagrams shown in Fig.~\ref{mefpiel}).
The most massive cluster is the Ara cluster (see also Sect.~4.1.) at ($\ell, b, v, \sigma$) $\approx$ 
($329.4\deg, -9.3\deg, 14722$ km s$^{-1}$, 1182 km s$^{-1}$).

\subsection{Pie diagrams}

An alternative view of the large-scale structures in this part of the sky ($240\deg \le \ell < 360\deg$,
$|\,b\,| < 30\deg$) is given by the pie diagrams in Fig.~\ref{mefpiel}. They show the galaxy distribution
out to 15\,000 km s$^{-1}$ (left panel), and 30\,000 km s$^{-1}$ (right panel).

In this representation, the Zone of Avoidance is almost indistinguisable from its neighbouring (unobscured)
region above and below the GP. In the Zone of Avoidance out to 30\,000 km s$^{-1}$ (middle right panel of Fig.~\ref{mefpiel}),
there are 1857 galaxies. In the adjacent region below the Galactic Plane (lower right panel), 1981 galaxies are present;
an almost identical number of galaxies. Above the GP more galaxies are present ($N = 3497$).
Note, however, that most of the redshifts in the Zone of Avoidance are for galaxies in the
Galactic latitude range of $|\,b\,| \sim 5\deg - 10\deg$. It is remarkable, nonetheless, that this region of the sky
that was previously devoid of information, now shows clear clusters, superclusters and voids. 

In the galaxy distribution out to 15\,000 km s$^{-1}$ (middle left panel), the Norma cluster at $(\ell, v)$ = ($325.3\deg, 4844$ km s$^{-1}$)
dominates the pie diagram. It is located centrally in an overdensity that connects the Centaurus cluster ($\ell, v)$ $\approx$ 
($302\deg, 3500$ km s$^{-1}$) and Hydra cluster ($\ell, v$) $\approx$ $(270\deg, 3800$ km s$^{-1}$) above the GP (upper left panel)
with the Pavo II cluster ($\ell, v$) $\approx$ ($332\deg, 4200$ km s$^{-1}$) below the GP (lower left panel). This entire large-scale
structure is the optical counterpart of the Great Attractor. The Centaurus-Crux cluster (almost dead centre in the middle left panel) at
$(\ell, v$) $\approx$ ($306\deg, 6200$ km s$^{-1}$) is also part of the GA.
The MEFOS observations of the Norma cluster form the basis for a detailed dynamical
study of this rich and nearby cluster at the heart of the Great Attractor. Recently, we have obtained
2dF spectroscopy of the Norma cluster (primarily for velocity dispersion measurements of the elliptical galaxies
in the cluster in order to determine the distance to the Norma cluster via the Fundamental Plane analysis). 
The results of this -- and the dynamical analysis of the cluster -- will be presented in a separate paper.

Out to 30\,000 km s$^{-1}$, many clusters in the velocity range 12\,500 -- 20\,000 km s$^{-1}$ are visible (middle right panel), the most prominent
is the Ara cluster at $(\ell, v$) = ($329\deg, 14722$ km s$^{-1}$).

\section{Summary}

Multi-fibre spectroscopy with MEFOS on the 3.6-m telescope of ESO has resulted in
764 redshifts in the southern Zone of Avoidance. Contrary to the SAAO and Parkes components of our
redshift survey in the Zone of Avoidance, MEFOS has focussed on the faint end of the galaxies in
our optical galaxy catalogue in the high galaxy-density regions. As a result, MEFOS has contributed 
significantly to the knowledge of low Galactic latitude clusters at $v \ga 12\,500$ km s$^{-1}$.

\acknowledgements

{This research has made use of 
the NASA/IPAC Extragalactic Database (NED), which is operated by the Jet 
Propulsion Laboratory, Caltech, under contract with the National Aeronautics 
and Space Administration. PAW kindly acknowledges financial support from the 
Antares fund and from the National Research Foundation. 
RCKK thanks CONACyT for their support (research grant 27602E). We thank the
Referee, Dr.~R.~Peletier, for the many useful comments and suggestions.
Part of this
survey was performed at the Kapteyn Astronomical Institute of the University of Groningen
and at the Observatoire de Paris-Meudon. Their support is greatfully acknowledged.}

\end{document}